\documentclass[proceedings, preprint]{rmaa}

\usepackage{paralist}

\usepackage{psfrag,color}

\SetYear{2013}
\SetConfTitle{MAGNETIC FIELDS IN THE UNIVERSE IV (2013)}

\title{Magnetic fields and halos in spiral galaxies}

\author{
  Marita Krause,\altaffilmark{1}}

\altaffiltext{1}{Max-Planck-Institut f\"ur Radioastronomie, Auf dem H\"ugel 69, 53121 Bonn, Germany
(mkrause@mpifr-bonn.mpg.de).}

\shortauthor{Krause}
\shorttitle{RevMexAA(SC)}

\listofauthors{M. Krause}
\indexauthor{Krause, M.}

\abstract{Radio continuum and polarization observations allow to reveal the magnetic field structure in the disk and
halo of nearby spiral galaxies, their magnetic field strength and vertical scale heights. The spiral galaxies studied
so far show a similar magnetic field pattern which is of spiral shape along the disk plane and X-shaped in the halo,
sometimes accompanied by strong vertical fields above and below the central region of the disk. The strength of the
halo field is comparable to that of the disk. While the total and turbulent magnetic field strength is (weakly)
increasing with the star formation, we could not find such a correlation for the ordered magnetic field strength. On
the contrary, there are indications that stronger star formation reduces the magnetic field regularity globally.\\
\
The magnetic field in spiral galaxies is generally thought to be amplified and maintained by dynamo action.
Investigation of the large- and small-scale magnetic fields during the galaxy's formation and cosmological evolution
lead to the picture that the turbulent dynamo amplifies the field strength to energy equipartition with the turbulent
(small-scale) gas, while the large-scale (mean-field) dynamo mainly orders the magnetic field. Hence, the large-scale
magnetic field pattern evolves with time. Supernova explosions causes a further continuous injection of turbulent
magnetic fields. Assuming that this small-scale field injection is situated only within the spiral arm region where
star formation mostly occurs lead to a large-scale field structure in which the magnetic field regularity is stronger
in the interarm region as observed in several nearby spiral galaxies, sometimes even forming magnetic arms.\\
\
For several spiral galaxies of different Hubble type and different star formation rates and efficiencies we found
similar scale heights of the total radio emission ($300 \pm 50$~pc for the thin disk and $1.8 \pm 0.2$~kpc for the
thick disk (halo)). This implies a relation between the galactic wind, the total magnetic field strength and the star
formation in the galaxy. A galactic wind may be essential for an effective dynamo action. Strong tidal interaction, however, seems to disturb the balance leading to deviating and locally different scale heights as observed in M82 and NGC~4631.
  }

\addkeyword{galaxies: spiral}
\addkeyword{galaxies: magnetic fields}
\addkeyword{galaxies: halos}
\addkeyword{galaxies: star formation}
\addkeyword{galaxies: evolution}
\addkeyword{radio continuum: galaxies}

\begin{document}
\maketitle

\section{Introduction}
\label{sec:intro}

The effects of magnetic fields on the physical processes in spiral galaxies, their disk-halo interaction and their
evolution have been frequently neglected in the past. Within the last 15 years, with increasing computing facilities,
some authors included them in their simulations of e.g. the interstellar medium and disk-halo interaction (e.g.
\citet{korpi+99}, \citet{avillez+05}) or in the evolution of spiral galaxies (e.g. \citet{pakmor+13}). Their result is
that magnetic fields play indeed an important role, even if the magnetic and cosmic ray energy density in the
interstellar medium is small compared to that of the rotation. The magnetic field energy density  is indeed comparable
to that of the turbulent gas motion and much higher than that of the thermal gas as has been determined for the nearby
galaxies NGC~6946 \citep{beck07} and M33 \citep{taba+08}. Hence, magnetic fields are dynamically important in the
processes of the interstellar medium. Direct comparison of 3-dimensional MHD simulations of an isolated galaxy with
and without a magnetic field show that the magnetic field lead to a lower star formation rate at later times, it
reduces the prominence of individual spiral arms and it causes weak outflows from the disk up to several kpc above and
below the disk \citep{pakmor+13}.\\
Observationally, the magnetic field in external galaxies can best be studied in the radio continuum emission in the cm
wavelength range. The total intensity of the synchrotron emission gives the strength of the total magnetic field. The
linearly polarized intensity reveals the strength and the structure of the resolved regular field in the sky plane
(i.e. perpendicular to the line of sight).  However, the observed polarization vectors suffer Faraday rotation and
depolarization (i.e. a decrease of the degree of linear polarization when compared to the intrinsic one) on the way
from the radiation's origin to us. Correction for Faraday rotation is possible with observations at different
wavelengths by determining the rotation measure RM (being proportional to $\int n_{\rm e} B_{\parallel}dl$ where
$n_{\rm e}$ is the thermal electron density and $B_{\parallel}$ the magnetic field strength parallel to the line of
sight l). The rotation measure itself can be used to correct the observed polarization angle and also to estimate the
strength of $B_{\parallel}$, its sign gives the direction of this magnetic field component. The field strength of both
components, parallel and perpendicular to the line of sight, together with the information of the intrinsic
polarization vectors enables us in principle to get a three-dimensional picture of the magnetic field.

\section{Faraday rotation and depolarization effects}
\label{sec:Faraday}

While the polarized intensity gives the {\bf orientation} of the magnetic field, the magnetic field {\bf direction}
can only be determined by the rotation measure. This implies that a large-scale regular (coherent) magnetic field can
only be deduced from the rotation measure pattern, while the polarized intensity may also originate from anisotropic
turbulent magnetic fields (e.g. compressed fields with opposite directions) in compressed or shocked regions.
As the polarization is only sensitive to the magnetic field orientation, the polarization angle can only be determined
with an $n \cdot \pi $ ambiguity. Further, depolarization effects have to be considered. We distinguish between
wavelength-independent and wavelength-dependent depolarization. The difference in depolarization at different
wavelengths in maps with the same linear resolution should be purely wavelength dependent where three different
wavelength-dependent depolarization effects are important to consider: the differential Faraday rotation, Faraday
dispersion, and a RM gradient within the beam \citep{burn66, sokoloff+98}. Faraday dispersion is due to turbulent
(random) magnetic fields within the source and between the source and us, whereas differential Faraday rotation and
depolarization by an RM gradient depends on the regular magnetic field within the emitting source. Especially
differential Faraday rotation may cause that the source is not transparent in polarization if the internal Faraday
rotation reaches values of $90\arcdeg$ or more which may be the case for observations of spiral galaxies seen edge-on
near the galactic midplane as e.g. in NGC~4631 \citep{mora+13} even in the decimeter wavelength-regime. The coming
polarization spectroscopy and RM-synthesis \citep{brentjens+05} will strongly reduce these effects.

\section{Magnetic field strength and star formation}
\label{sec:strength}
\begin{figure}
\centering
\includegraphics[bb = 11 42 545 608, width=\columnwidth]{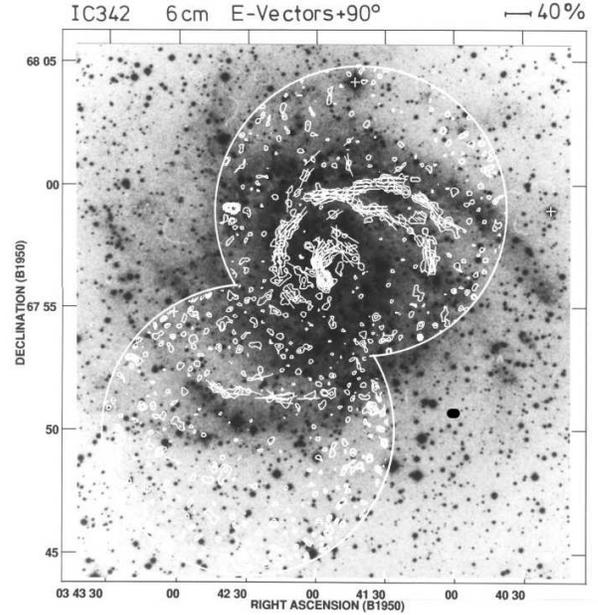}
\caption{Radio continuum emission of the spiral galaxy IC342 at $\lambda6.2$cm (4.8~GHz) from VLA observations with a
resolution of $16 \arcsec~HPBW$, overlaid on a POSS optical image. The contours give the polarized intensities, the
vectors the apparent magnetic field orientation with their lengths being proportional to the degree of polarization
\citep{krause93}.
}
\label{ic342}
\end{figure}

The total magnetic field strength in a galaxy can be estimated from the nonthermal radio emission under the assumption
of equipartition between the energies of the magnetic field and the relativistic particles (the so-called {\em energy
equipartition}) as described in \citet{beck+05}. The mean equipartition value for the total magnetic field strength
for a sample of 74 spiral galaxies observed by \citet{niklas95} is on average $9\,\pm 3\,\mu$G but reaches locally
higher values {\em within} the spiral arms of up to $20 - 25\,\mu$G in M51 \citep{fletcher+2011}.  The strength of the
{\bf ordered}  magnetic fields in spiral galaxies are typically 1--5~$\mu$G, and may reach locally values up to $10 -
15\,\mu$G as e.g. in NGC~6946 \citep{beck07} and M51 \citep{fletcher+2011}. The field strengths in the halo are
comparable to the those in the disk (see Sect.~\ref{sec:structure}).\\
\
\begin{figure}
\centering
\includegraphics[bb = 14 14 132 159, width=\columnwidth]{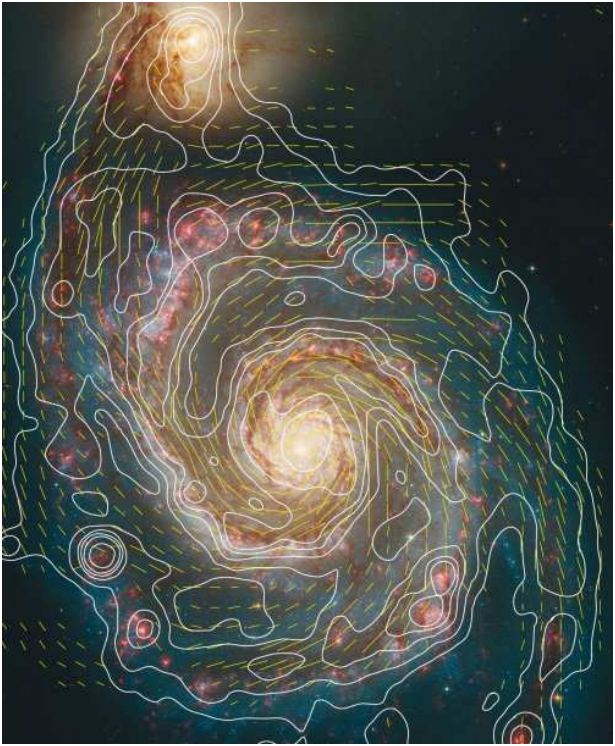}
\caption{Radio continuum emission of the spiral galaxy M51 at $\lambda6.2$cm (4.8~GHz) from VLA and 100-m Effelsberg
observations with a resolution of $15 \arcsec~HPBW$, overlaid on a Hubble Space Telescope optical image [image credit:
NASA, ESA, S. Beckwith (STScI) and The Hubble Heritage Team (STScI/AURA)]. The contours give the total intensities,
the vectors the apparent magnetic field orientation (i.e. not corrected for Faraday rotation) with their lengths being
proportional to the polarized intensity \citep{fletcher+2011}.
}
\label{m51}
\end{figure}
The {\bf turbulent} magnetic field is typically strongest within the optical spiral arms, whereas the regular fields
are strongest in between the optical spiral arms, or at the inner edge of the density-wave spiral arm as seen in M51
\citep{fletcher+2011} (Figure~\ref{m51}). Sometimes, the interarm region is filled smoothly with regular fields, in
other cases the large-scale field form long filaments of polarized intensity like in IC342 (Figure~\ref{ic342})
\citep{krause93} or so-called {\em magnetic spiral arms} like in NGC~6946 \citep{beck+96}.\\
\
\
\
Strongly interacting galaxies or galaxies with a high star formation rate (SFR) tend to have generally stronger total
magnetic fields. The latter fits to the equipartition model for the radio-FIR correlation \citet{niklas+97}, according
to which the nonthermal emission increases $\propto SFR^{1.3 \pm 0.2}$ and the \emph{total}, mostly turbulent
magnetic field strength $ \rm B_t$ increases $\propto SFR^{0.34 \pm 0.14}$.\\
\
\
No similar simple relation is known for the \emph{ordered} magnetic field strength. We integrated the polarization
properties in 41 nearby spiral galaxies and found that (independent of inclination effects) the degree of linear
polarization is lower ($ < 4\%$) for more luminous galaxies, in particular those for $ L_{4.8} > 2 \times
10^{21}~\rm{W Hz^{-1}}$  \citet{stil+09}. The radio-brightest galaxies are those with the highest SFR. Though dynamo
action needs star formation and supernova remnants as the driving force for velocities in vertical direction, we
conclude from our observations that stronger star formation reduces the magnetic field regularity. On kpc-scales,
\citet{chyzy08} analyzed the correlation between magnetic field regularity and SFR locally within one galaxy,
NGC~4254. While he found that the total and random field strength increase locally with SFR, the ordered field
strength is locally uncorrelated with SFR.\\
\section{Magnetic field structure in spiral galaxies}
\label{sec:structure}
\begin{figure*}[!t]
  \includegraphics[bb=28 185 564 684, width=\columnwidth]{n5775.tp6_new.ps}%
  \hspace*{1.3\columnsep}%
  \includegraphics[bb=135 125 465 475, width=0.86\columnwidth]{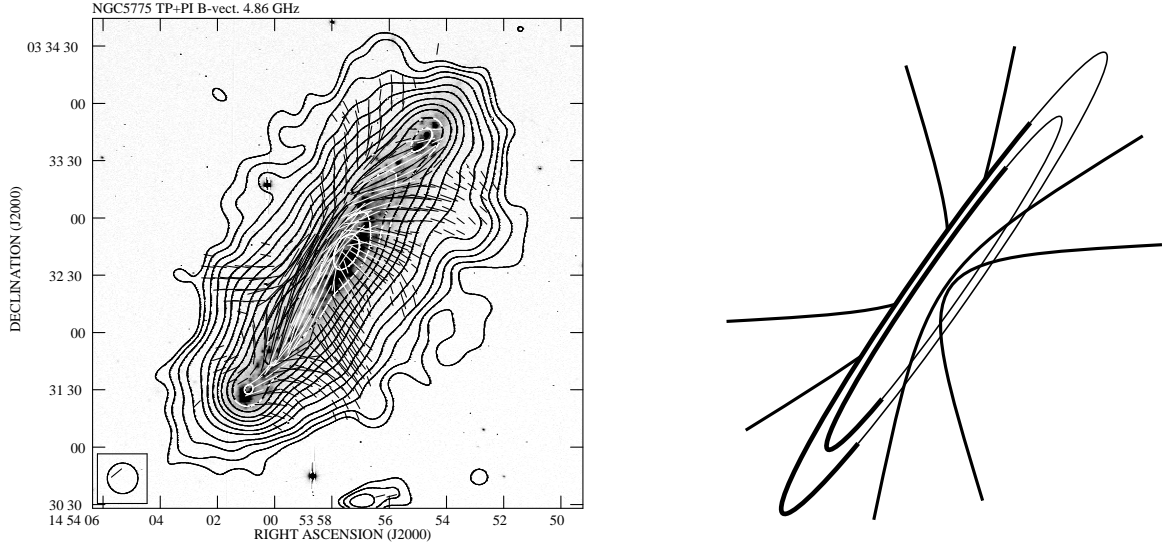}
  \caption{Total intensity contours map with apparent magnetic field orientation at 4.8~GHz of NGC~5775 (from the VLA
  with a resolution of $16 \arcsec~HPBW$) overlaid on an H$\alpha$ image (left) and sketch of the regular magnetic
  field configuration in the disk and in the halo (right) (from \citealp{soida+2011}).
  }
  \label{n5775}
\end{figure*}

\begin{figure*}
\centering
\includegraphics[bb = 14 14 373 218, width=0.95\textwidth]{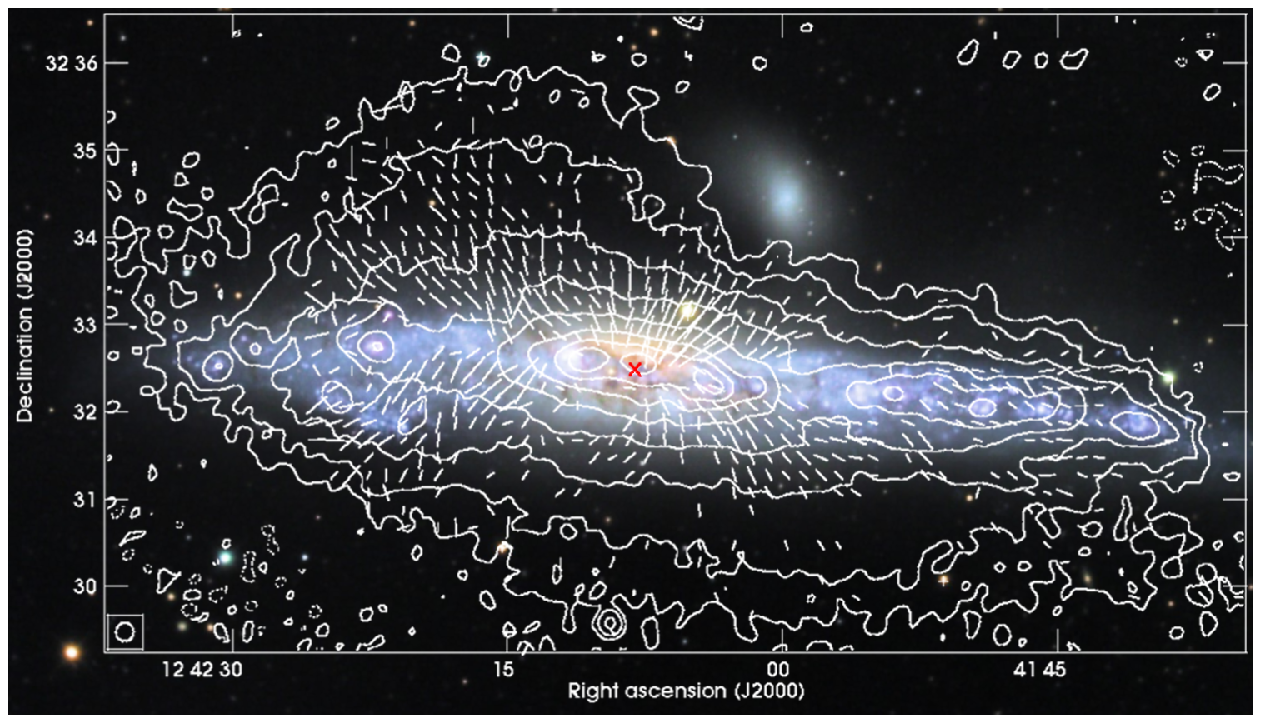}
\caption{Radio continuum emission of the edge-on spiral galaxy NGC~4631 at $\lambda6.2$cm (8.4~GHz) from VLA and 100-m
Effelsberg observations with a resolution of $12 \arcsec~HPBW$, overlaid on a colour-scale optical DSS image. The
contours give the total intensities, the vectors the apparent magnetic field orientation with their lengths
proportional to the polarized intensity \citep{mora+13}.
}
\label{n4631}
\end{figure*}

\
Observations of spiral galaxies seen face-on reveal a large-scale magnetic field pattern along the plane of the
galaxy. The magnetic field lines generally follow a spiral structure with pitch angles from $10\arcdeg$ to $40\arcdeg$
which are similar to the pitch angles of the optical spiral arms, as visible e.g. in (Figure~\ref{m51}). This
large-scale pattern is accompanied by a small-scale magnetic field which is stronger within the optical spiral arms.
The magnetic field is thought to be amplified and maintained by dynamo action, especially the large-scale structure by
the action of the mean-field dynamo \citep{ruzmaikin+88} which predicts an axisymmetric mode (ASS) along the galactic
plane of the galaxy to be excited most easily. The mean-field dynamo theory alone, however, cannot explain why the
strength of the large-scale magnetic field is higher in the interarm region as discussed in Sect.~\ref{sec:strength}.
This will further be discussed in Sect.~\ref{sec:evolution}.  The mean field theory also cannot explain why the
magnetic pitch angles are similar to the pitch angles of the optical spiral arms (as summarised in
\citealp{fletcher10}). Within the dynamo theory the magnetic pitch angles are simply determined by the ratio of the
radial to the azimuthal magnetic field components ($p = \arctan B_r / B_\varphi = - \sqrt{ R_\alpha / R_\omega}$, see
also Sect.~\ref{sec:evolution}) \citep{ruzmaikin+88}, whereas the spiral arm pitch angle may even be related to the
supermassive black hole mass \citep{seigar+08}.\\
Observation of spiral galaxies seen edge-on show in general a plane-parallel magnetic field structure along the
midplane which is the expected projection of the spiral field in the disk as observed in face-on galaxies. This is
also the case in NGC~4631 as detected by \citet{mora+13}.\\
In the halo the ordered magnetic field is X-shaped as indicated in the sketch for NGC~5775 observed with an
inclination i$=86 \arcdeg$ (Figure~\ref{n5775}). In some galaxies the X-shaped halo field is accompanied by strong
vertical components above and/or below the central region as in NGC~5775 (Figure~\ref{n5775}) and NGC~4631
(Figure~\ref{n4631}). Reliable RM values of the X-shaped field in the halo are still missing, hence we cannot decide
observationally whether these are regular or anisotropic turbulent fields. They can also be a mixture of both.\\
\
\
\
\
\
In general the strength of the ordered halo field is comparable to the strength of the large-scale disk field. It
cannot be explained by the classical mean-field dynamo operating in the disk. Even though this is also accompanied by
a poloidal halo field, this is by a factor of about 10 weaker than the observed halo field. Either there is also dynamo
action in the halo or a galactic wind is needed to transport magnetic field from the disk into the halo. This will be
further discussed in Sect.~\ref{sec:scale height}\\
\
\
\section{Dynamo action and the evolution of the large-scale magnetic field}
\label{sec:evolution}

\begin{figure}
\centering
\includegraphics[bb = 97 31 578 707,angle=-90,width=\columnwidth]{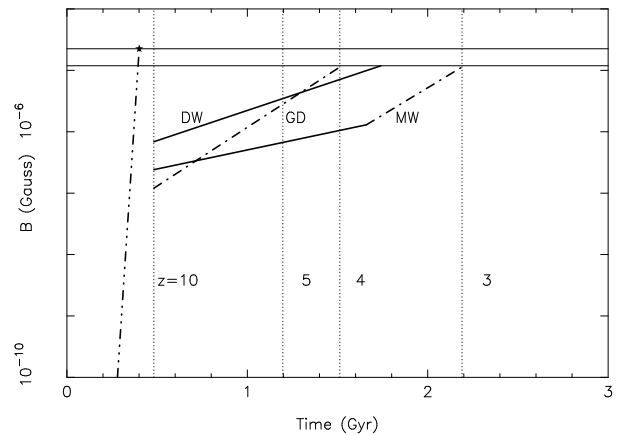}
\caption{Evolution of magnetic field strength in dwarf galaxies (DG), Milky-Way-type galaxies (MW), and in giant disk
galaxies (GD). The thick dashed-dot-dot-line shows the evolution of the small-scale magnetic field generated by the
small-scale dynamo. The evolution of the large-scale magnetic field generated by the mean-field dynamo in
quasi-spherical galaxies are shown by the thick solid line and that in thin-disk galaxies as thick dashed-dot
dashed-lines \citep{arshakian+09}.
}
\label{evolution}
\end{figure}

\begin{figure}
\centering
\includegraphics[bb = 14 14 497 506,width=\columnwidth]{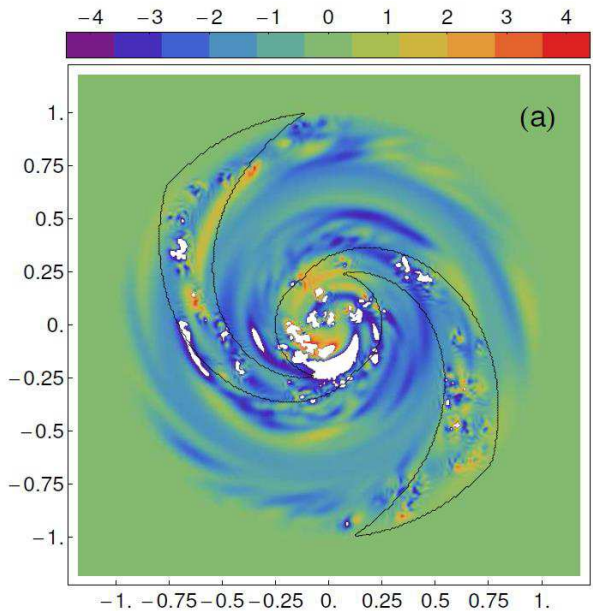}
\caption{Colour coded image of the modeled azimuthal magnetic field strength in the disk of a spiral galaxy with
turbulent magnetic field injection assumed within the normal spiral arm regions as indicated by the thin black lines
after 13.2~Gyr. The field strength is given in $\mu$G \citep{moss+13}.
}
\label{arms}
\end{figure}

The large-scale magnetic field can only be amplified and maintained by dynamo action. While a large-scale dynamo is
necessary to produce a large-scale magnetic field structure, field amplification alone is faster by the action of a
small-scale dynamo \citep{beck+94,beck+96}. For a galactic disk, the large-scale dynamo is the $\alpha\Omega$-dynamo,
simplified by the mean-field dynamo equations. Solution of these equations are the large-scale dynamo-modes, with the
axisymmetric spiral field structure (ASS) being the dominant mode (m=0) which is generated easiest, followed by the
bisymmetric structure (BSS, m=1) and higher modes. The 3-dimensional field configurations can be either symmetric (of
quadrupole type) or asymmetric (of dipole type) with respect to the galactic plane, where the poloidal field component
is about a factor of 10 weaker than the disk field. According to the dynamo theory the pitch angle of the magnetic
field spiral is determined by the dynamo numbers $R_\alpha$ and $R_\omega$, not by the pitch angle of the gaseous
spiral arms. The modes determined observationally in a dozen of nearby galaxies and their relative amplitudes are
summarised by \citet{fletcher10}. The dominating mode in the disk is indeed the ASS.\\
As part of the SKA design study we investigated the large- and small-scale dynamo action and the ordering process of
the large-scale magnetic field structure during galaxy formation and cosmological evolution \citep{arshakian+09}.
Turbulence generated in protogalactic halos by thermal virialization can drive an effective turbulent (small-scale)
dynamo which amplifies the field strength to energy equipartition with the turbulent gas (beginning at $z \approx 10 $ in
Figure~\ref{evolution}). The large-scale dynamo mainly orders the magnetic field with timescales determined by the
mean-field dynamo theory. Hence, the large-scale fields evolve with time. Galaxies similar to e.g. the Milky Way
formed their disks at $z \approx 10$. Regular fields of $\mu$G strength and a few kpc coherence length were generated
within 2 Gyr (beginning at $z \approx 20$ in Figure~\ref{evolution}), but field ordering on the coherence length of
the galaxy size requires additional 6~Gyr for Milky-Way-type galaxies. Dwarf galaxies can already host fully coherent
fields at $z \approx 1$ while giant disk galaxies may not have reached fully coherent filed pattern in the Universe'
lifetime up to now \citep{arshakian+09}.\\
\
\
\
\
\citet{arshakian+11} studied the field ordering of a so-called ``spotty'' magnetic field structure in more detail
assuming that the large-scale dynamo starts from coherent fields in spots of 100~pc in size and 0.02~$\mu$G in
strength. The evolution of these magnetic spots is simulated in a model. A star formation in a galaxy causes -via
supernova explosions- a continuous injection of turbulent magnetic fields. \citet{moss+12} combined the interaction of
magnetic fields generated by small-scale dynamo action in discrete star formation regions together with the mean-field
dynamo action. Assuming that the injection of small-scale fields is situated only within the gaseous spiral arm
regions where star formation mostly occurs, \citet{moss+13} obtained field structures with magnetic arms located between the
spiral arms as discussed in Sect.~\ref{sec:strength} (see Figure~\ref{arms}).
\
\
\

\section{Vertical scale heights and galactic wind}
\label{sec:scale height}

We determined the vertical scale heights of the total power radio emission in several edge-on spiral galaxies. The
results are summarized in Tab.~\ref{tableone}. For all galaxies except M104 (NGC4594) a (two-component) exponential
fit was better than a gaussian fit. For the first 5 galaxies in Tab.~\ref{tableone} and for NGC~4631 single-dish
(100-m Effelsberg) and interferometer (VLA) data were merged in order not to miss extended flux because of missing
spacings. The scale heights were determined from $\lambda6$~cm observations except for NGC~5907 (observed at
$\lambda20$~cm). At $\lambda6$~cm the vertical {\em scale heights} of the thin disk and the thick disk/halo in the
sample of the first five galaxies in Tab.~\ref{tableone} are similar and have a mean value of $300 \pm 50$~pc for the
thin disk and $1.8 \pm 0.2$~kpc for the thick disk. This sample of galaxies include the one with the brightest halo
observed so far, NGC~253, with strong star formation, as well as one of the weakest halos, NGC~4565, with weak star
formation. The SFR-values were determined directly from the IR-emission given by \citet{young+98} according to
\citet{kennicut98}, the SFE-values in Tab.~\ref{tableone} were determined from values for the molecular masses given
in the literature.\\
If synchrotron emission is the dominant loss process of the relativistic electrons the outer shape of the radio
emission should be dumbbell-like as the local scale height depends on the local magnetic field strength. In fact, a
dumbbell shape of the total radio intensity  has been observed in several edge-on galaxies like e.g. NGC~253
\citep{heesen+09a} and NGC~4565. As the synchrotron lifetime $t_{syn}$ at a fixed frequency is proportional to the
total magnetic field strength $B_t^{-1.5}$, a cosmic ray bulk speed (velocity of a galactic wind) can be defined as
$v_{CR} = h_{CR}/t_{syn} = 2 h_z/t_{syn}$ in the case of equipartition, where $h_{CR}$ and $h_z$ are the scale heights
of the cosmic rays and the observed radio emission at this frequency. For NGC~253 \citet{heesen+09a} determined the
cosmic ray velocity to $300 \pm 30$~km/s in the north-eastern halo. As this is similar to the escape velocity, it
shows the presence of a galactic wind in NGC~253. Further, the similarity of the observed radio scale heights suggest
a self regulation mechanism between the galactic wind velocity, the total magnetic field strength and the star
formation rate SFR in the disk: $v_{CR}\propto B_t^{1.5} \propto SFR^{\approx 0.5}$ where the relation between $B_t$
and SFR refer to the equipartion model for the radio-FIR relation \citep{niklas+97}.\\
\
\
\begin{table*}[ht]
\setlength{\tabnotewidth}{\textwidth}
\tablecols{9}
 \caption{Vertical scale heights, star formation rates (SFR) and efficiencies (SFE), averaged total magnetic field
 strengths $\rm B_t$, inclination i and Hubble-type for our sampe of edge-on spiral galaxies\tabnotemark{a}}
  \begin{tabular}[]{lcccccccc}
  \toprule
          &\multicolumn{2}{c}{\textbf{Vertical scale heights}} & & & & & &\\
  \cmidrule{2-3}
   galaxy & thin disk & thick disk & \textbf{SFR}(IR) & \textbf{SFE} &  $\mathbf {B_t}$ & i & Hubble & References \\
          & [pc] & [kpc] & [$ \rm{M}_\odot \rm{yr}^{-1} $] & [$ \rm{L}_\odot \rm{M}_\odot^{-1} $] & [$\mu$G] &
          [$\arcdeg $]   & type & for scale heights\\
   \hline
   NGC253 & $380 \pm 60$  & $1.7 \pm 0.1$ & 6.3 & 14 & 12 & 78 & Sc & \citealp{heesen+09a}\\
   NGC891 & 270 & 1.8 & 3.3 & 5.0 & 6 & 88 & Sb & \citealp{dumke+98}\\
   NGC3628 & 300 & 1.8 & 1.1 & 4.9 & 6 & 89 & Sb pec & \citealp{dumke+98}\\
   NGC4565 & 280 & 1.7 & 1.3 & 3.2 & 7 & 86 & Sb & \citealp{dumke+98}\\
   NGC5775 & $240 \pm 30$ & $2.0 \pm 0.2$ & 7.3 & 6.1 & 8 & 86 & Sbc & \citealp{soida+2011}\\

   \hline
   mean    & $\mathbf {300 \pm 50}$ & $\mathbf {1.8 \pm 0.2}$\\
   \hline
   NGC5907 & 340 & $\ge 1.5$ & 1.3 & 4.0 & 5 & 87 & Sc & \citealp{dumke+00}\\
   M104    &\multicolumn{2}{c}{$1.4 \pm 0.2$ gaussian}& 1.2 & 4.2 & 4 & 84 & Sa & \citealp{krause+06}\\
   NGC4631 & $370 \pm 280$ & $2.3 \pm 0.9$ & 2.1 & 9.9 & 6 & 86 & SBd & \citealp{mora+13}\\
   M82     &\multicolumn{2}{c}{see text in Sect.~\ref{sec:scale height}} & 1.8 & 22  &35 & 79 & SBc &
   \citealp{adebahr+13}\\
  \bottomrule
  \tabnotetext{a}{The vertical scale heights are determined from 4.8~GHz data except for NGC~5907 (at 1.4~GHz). The
  4.8~GHz observations for the first five galaxies and NGC~4631 are merged single-dish (100-m Effelsberg) and
  interferometer (VLA) data.}
 \end{tabular}
 \label{tableone}
\end{table*}
\
The scale heights in NGC~5907 are measured from $\lambda20$~cm observations taken with the VLA \citep{dumke+00}. They
are similar to the values of the sample of 5 galaxies discussed above. If synchrotron losses dominate in this galaxy,
its scale heights are expected to be somewhat larger than those at $\lambda6$~cm, however the maps may suffer large
extended structure \citep{dumke+00}.\\
The Sombrero galaxy M104 is classified as an Sa galaxy and shows a huge bulge with an elliptical mass distribution.
The expected z-distribution of a relatively thin layer (the disk) inside a nearly spherical gravitational potential is
in fact a Gaussian \citep{combes91}. The bulge in M104 may be due to a dissolving bar as proposed by Emsellem,
1995.\\
Recent determination of the scale heights in M82 and NGC~4631 also yielded different values: the scale heights in M82
are by a factor of $\approx 3$ smaller than the mean values mentioned above. The scale heights in the north are larger
than those south of the disk \citep{adebahr+13}. The scale heights for both, the thin and thick disk in NGC~4631 vary
strongly within the galaxy being significantly larger in some areas than the mean values in Tab.~\ref{tableone}
\citep{mora+13}. Both galaxies show -different to the first five galaxies in Tab.~\ref{tableone}- strong signs of
tidal interaction like HI tails and bridges \citep{yun+93,rand94} where M82 might even have lost its outer HI disk by
tidal disruption. Hence, from present observations of edge-on galaxies we conclude that while star formation and even
starbursts in the disk alone do not significantly change the scale heights of the disk and halo, (strong) tidal
interactions may well modify these parameters.\\
The observations of similar scale heights for not strongly interacting galaxies (the first five galaxies in
Tab.~\ref{tableone} and probably NGC~5907 as well) suggest the existence of a galactic wind in all of them. It may be
essential for the formation of large-scale magnetic field in the halo as discussed in Sect.~\ref{sec:structure}.\\
Indeed, model calculations of the mean-field $\alpha\Omega$-dynamo for a disk surrounded by a spherical halo including
a galactic wind \citep{brandenburg+93, moss+10} simulated similar magnetic field configurations to the observed ones.
Meanwhile, MHD simulations of disk galaxies including a galactic wind implicitly may explain the X-shaped field
\citep{gressel+08, hanasz+09a}. The first global, galactic scale MHD simulations of a CR-driven dynamo give promising
results resembling the observations and show directly that small scale magnetic flux is transported from the disk into
the halo \citep{hanasz+09b}. A galactic wind can also solve the helicity problem of dynamo action (e.g.
\citealp{sur+07}). Hence, a galactic wind may be essential for an effective dynamo action and the observed X-shaped
magnetic field structure in edge-on galaxies.\\

\end{document}